\documentclass[twocolumn,showpacs,superscriptaddress]{revtex4-1}
\usepackage{graphicx}
\usepackage{units}
\usepackage{multirow}
\usepackage{bigstrut}
\usepackage{overpic}
\usepackage{array}
\usepackage{color}
\usepackage{amsmath}
\RequirePackage{lineno}

\usepackage[perpage,symbol]{footmisc}

\newcommand{\etap}{\eta^{\prime}}
\newcommand{\pn}{p\bar{n}}

\newcommand{\ee}{e^+e^-}

\newcommand{\ds}{D_s}
\newcommand{\dss}{D^*_s}
\newcommand{\dsp}{D^+_s}
\newcommand{\dsm}{D^-_s}
\newcommand{\dssp}{D^{*+}_s}
\newcommand{\dssm}{D^{*-}_s}

\newcommand{\mevc}{\,\unit{MeV}/c}

\newcommand{\gev}{\,\unit{GeV}}

\newcommand{\gevcc}{\,\unit{GeV}/c^2}

\newcommand{\br}[1]{\mathcal{B}_{#1}}
\newcommand{\ks}{K^0_S}

\newcommand{\mrec}{M_\mathrm{rec}}
\newcommand{\mmiss}{M_\mathrm{miss}}

\newcommand {\result}   {(1.21\pm0.10\pm0.05)\times10^{-3}}

\newcommand {\dspn} {D^+_s\rightarrow\pn}
\newcommand {\ecm} {E_\mathrm{cm}}
\newcommand {\modea} {\ks K^-}
\newcommand {\modeb} {K^- K^+\pi^-}
\newcommand {\modec} {\ks K^-\pi^0}
\newcommand {\moded} {K^- K^+\pi^-\pi^0}
\newcommand {\modee} {\ks K^+\pi^-\pi^-}
\newcommand {\modef} {\pi^-\pi^+\pi^-}
\newcommand {\modeg} {\pi^-\eta}
\newcommand {\modeh} {\rho^-\eta}
\newcommand {\modei} {\pi^-\etap}
\newcommand {\modej} {\pi^-\etap}
\newcommand {\modek} {K^- \pi^+\pi^-}

\begin{document}

%\linenumbers

\title{\boldmath Observation of $\dspn$ and confirmation of its large branching fraction}

\author{
\small
\begin{center}
M.~Ablikim$^{1}$, M.~N.~Achasov$^{9,d}$, S. ~Ahmed$^{14}$, M.~Albrecht$^{4}$, M.~Alekseev$^{55A,55C}$, A.~Amoroso$^{55A,55C}$, F.~F.~An$^{1}$, Q.~An$^{52,42}$, J.~Z.~Bai$^{1}$, Y.~Bai$^{41}$
% ...
\\
\vspace{0.2cm}
(BESIII Collaboration)\\
\vspace{0.2cm} {\it
$^{1}$ Institute of High Energy Physics, Beijing 100049, People's Republic of China\\
% ...
$^{4}$ Bochum Ruhr-University, D-44780 Bochum, Germany\\
% ...
$^{9}$ G.I. Budker Institute of Nuclear Physics SB RAS (BINP), Novosibirsk 630090, Russia\\
% ...
$^{14}$ Helmholtz Institute Mainz, Johann-Joachim-Becher-Weg 45, D-55099 Mainz, Germany\\
% ...
$^{41}$ Southeast University, Nanjing 211100, People's Republic of China\\
$^{42}$ State Key Laboratory of Particle Detection and Electronics, Beijing 100049, Hefei 230026, People's Republic of China\\
% ...
$^{52}$ University of Science and Technology of China, Hefei 230026, People's Republic of China\\
% ...
$^{55}$ (A)University of Turin, I-10125, Turin, Italy; (B)University of Eastern Piedmont, I-15121, Alessandria, Italy; (C)INFN, I-10125, Turin, Italy\\
% ...
$^{d}$ Also at the Novosibirsk State University, Novosibirsk, 630090, Russia\\
% ...
}
\end{center}
\vspace{0.4cm}
}
\noaffiliation{}
%%%%%%%%%%%%%%%%%%%%%%%%%%%%%%%%%%%%%%%%%%%%%%%%%%%%%%%%%%%%%%%%%%%%%%%%%%%%%%%%%%%%%%%%%%

\begin{abstract}

The baryonic decay $\dspn$ is observed %with a statistical significance of much larger than 10 standard deviations
, and the corresponding branching fraction is measured to be $\result$, where the first uncertainty is statistical and second systematic. The data sample used in this analysis was collected with the BESIII detector operating at the BEPCII $e^+e^-$ double-ring collider with a center-of-mass energy of 4.178~GeV and an integrated luminosity of 3.19~fb$^{-1}$. The result confirms the previous measurement by the CLEO Collaboration and is of greatly improved precision. This result will improve our understanding of the dynamical enhancement of the W-annihilation topology in the charmed meson decays.

\end{abstract}

\pacs{13.20.Fc, 12.38.Qk, 14.40.Lb}

\maketitle

%%%%%%%%%%%%%%%%%%%%%%%%%%%%%%%%%%%%%%%%%%%%%%%%%%%%%%%%%%%%%%%%
%%%%%     Introduction       Part                  %%%%%%%%%%%%%
%%%%%%%%%%%%%%%%%%%%%%%%%%%%%%%%%%%%%%%%%%%%%%%%%%%%%%%%%%%%%%%%
%\begin{multicols}{2}

%\section{Introduction}\label{sec:intro}

The decay $\dspn$ is the only kinematically allowed baryonic decay of the three ground-state charmed mesons $D^0$, $D^+$, and $D^+_s$.  It provides a unique probe of hadronic dynamics and is of great importance to the study of weak annihilation decays of charmed mesons~\cite{Gershtein:1976aq,Pham:1980,Bediaga:1992, Chen:2008pf,Bigi:1992}.
At the short-distance level,
under the vacuum-insertion approximation, its branching fraction (BF) is predicted to be very small (of the order $10^{-6}$) owing to chiral suppression by the factor $(m_\pi/m_{\ds})^4$ which follows from the partially conserved axial current~\cite{Chen:2008pf}. This physically corresponds to the mechanism of helicity suppression.

The CLEO Collaboration reported evidence for the decay $\dspn$ with $13.0\pm3.6$ signal events, resulting in an anomalously large BF of $\br{}(\dspn)=(1.30\pm0.36^{+0.12}_{-0.16})\times10^{-3}$~\cite{Cleo-c:pn}.
This unexpectedly large BF stimulates the interest of theorists. Many phenomenological possibilities have been proposed to explain the apparent discrepancy between theoretical predictions and the experimental measurement, $e.g.$, the not well justified factorization ansatz due to the light mass of charm quark and the complicated final state interaction at the threshold of $\pn$ production~\cite{Pham:1980,Bediaga:1992,Chen:2008pf}, a contribution of additional decay mechanisms such as final state scattering~\cite{Chen:2008pf}, or the effect of the time-like baryonic form factors from the axial vector currents~\cite{Hsiao:2015}.
Experimentally, the confirmation of the observation of the decay $\dspn$ by different experiments is highly desirable, and a much improved precision on its decay BF is necessary to distinguish between different phenomenological models and understand the decay dynamics of charmed mesons. The $\ee$ annihilation sample collected at $\sqrt{s}=4.178$\,$\gev$ with the Beijing Spectrometer (BESIII) in 2016, which corresponds to  an integrated luminosity of 3.19~fb$^{-1}$ and is 5 times larger in statistics compared to the CLEO data, provides a good opportunity for this measurement.

BESIII is a general-purpose
detector with 93\% coverage of the full solid angle.
Details of the detector can be found in Ref.~\cite{:2009vd}.
In 2015, BESIII was upgraded by replacing the two endcap time-of-flight (TOF) systems with a new detectors that use multi-gap resistive plate chambers (MRPC), which achieve a time resolution of 60\,ps~\cite{Wang:2016bzv}.

A {\sc geant4}-based~\cite{geant4} Monte Carlo (MC) simulation software package, which includes the description of the BESIII detector geometry and its response, is used to generate MC simulated event samples.
The simulation includes the beam energy spread and initial state radiation (ISR) in the $\ee$ annihilations modeled with the generator {\sc ConExc}~\cite{Ping:2013jka}.
The final state radiation (FSR) from charged tracks is incorporated with the {\sc photos} package~\cite{photos}.
The generic MC samples, consisting of the production of open charm processes, the ISR return to low-mass charmonium ($\psi$) states, and continuum processes (quantum electrodynamics processes and continuum production of light quarks $q\bar{q}$, $q=u,d,s$), have a size corresponding to an integrated luminosity 35 times larger than that of the data.
The known particle decays are generated using {\sc evtgen}~\cite{evtgen} with the BFs taken from
the Particle Data Group (PDG)~\cite{PDG}, and the remaining unknown decays of low mass $\psi$ states are generated with {\sc lundcharm}~\cite{lund}.
We also generate a signal MC sample of $4\times10^6$ $\ee\to\dssp \dsm\to\dsp\gamma\dsm$ events, in which $\dsp$ decays to $p\bar{n}$, while the $\dsm$ is set to decay generically, while the charge conjugated modes is also included; this sample is used to obtain the shapes of kinematic variables in signal decays and to estimate systematic uncertainties.
Throughout the Letter, charge conjugated modes are implicitly implied, unless otherwise noted.

%\section{Data Analysis}

In this analysis, the $\dsp$ sample is predominantly produced in the reaction $\ee\rightarrow\dssp\dsm\rightarrow\dsp\gamma\dsm$.
We fully reconstruct a $\dsm$ meson, named ``single tag (ST)'', in eleven decay modes that correspond to 25\%~of the total decay width \cite{PDG}:
$\modea$, $\modeb$, $\modec$, $\moded$, $\modee$, $\modef$, $\modeg$,
$\modeh$, $\modei$ (with $\etap\to\pi^+\pi^-\eta$), $\modej$ (with $\etap\to\gamma\pi^+\pi^-$) and $\modek$.
Then in the ST $\dsm$ sample, we further require an isolated photon consistent with $D^{*+}_{s}$ decay and reconstruct the
$\dspn$ signal in the side recoiling against the $\dsm$ candidate, referred to as the ``double tag (DT)''.
Both the $\dsp$ directly produced in the $\ee$ annihilation and the one from $\dssp$ decay are considered.
Thus, the numbers of ST ($N_{\rm ST}^i$) and DT ($N_{\rm DT}^i$) candidates for a specific tag mode $i$ are
%{\small
\begin{eqnarray}
N_{\rm ST}^i &=& 2N_{\rm tot}\cdot\br{i}\cdot\epsilon_{\rm ST}^i,\label{1}\\
N_{\rm DT}^i &=& 2N_{\rm tot}\cdot\br{i}\cdot\br{\dssp\to\gamma \dsp}\cdot\br{\dspn}\cdot\epsilon_{\rm DT}^i,\label{2}
\end{eqnarray}%}
\noindent
where $N_{\rm tot}$ is the total number of $\ee\to\dssp\dsm+c.c.$ events in the data,
$\br{i}$, $\br{\dssp\to\gamma\dsp}$, and $\br{\dspn}$ are the BFs for $\dsm$ tag mode $i$, $\dssp\to\gamma\dsp$, and $\dspn$, respectively,
and $\epsilon_{\rm ST(DT)}^i$ is the ST\,(DT) detection efficiency.
The factor 2 indicates that the signal $\dsp$ is either directly produced in the $\ee$ annihilation or from $\dssp$ decay.
Based on Eqs.~(\ref{1}) and (\ref{2}), combining the eleven ST modes leads to the expression
\begin{eqnarray}
    \br{D_s^+\to p\bar{n}}
    &=&\frac{1}{\br{D^*_s\to\gamma D_s}}\cdot\frac{N_{\rm DT}^{\rm tot}}{\sum_{i} N_{\rm ST}^i\cdot\epsilon_{\rm DT}^i/{\epsilon_{\rm ST}^i}},\label{3}
    \label{equation3}
\end{eqnarray}
where $N_{\rm DT}^{\rm tot}$ is the total number of DT signal events reconstructed from all ST modes.

All charged tracks are reconstructed from hits in the main drift chamber (MDC) with a polar angle $\theta$ (with respect to the beam direction) within $|\cos\theta|<0.93$.
Charged tracks, except for those from $\ks$ decays, are required to have a point of closest approach to the interaction point (IP) within $\pm10$\,cm along the beam direction and within $1$\,cm in the plane perpendicular to the beam axis.
Particle identification (PID) is performed by combining the specific energy loss d$E$/d$x$ measured in the MDC and the TOF information.
A charged $\pi$($K$) candidate is identified by requiring the PID likelihood value $\mathcal{L}(\pi)>\mathcal{L}(K)$, $\mathcal{L}(\pi)>0$ ($\mathcal{L}(K)>\mathcal{L}(\pi)$, $\mathcal{L}(K)>0$).

Photon candidates are reconstructed with energy deposits in the electromagnetic calorimeter (EMC) that are not associated with reconstructed charged tracks.
The photon is required to have an energy larger than 25\,MeV in the barrel region ($|\cos\theta|<0.8$), or 50\,MeV in the endcap region ($0.86<|\cos\theta|<0.92$).
To suppress electronic noise and energy deposits unrelated to the events, the shower time in the EMC must be within 700\,ns of the event start time~\cite{t0}.
The $\pi^0$ and $\eta$ candidates are reconstructed from  $\gamma\gamma$ pairs with an invariant mass $M_{\gamma\gamma}$ within (0.115, 0.150)\,$\gevcc$ and (0.50, 0.57)\,$\gevcc$, respectively.
Candidates with both photons in the endcap regions are rejected due to the bad energy resolution.
To improve the momentum resolution, a 1C kinematic fit is performed, constraining $M_{\gamma\gamma}$ to the nominal $\pi^0$ or $\eta$ mass~\cite{PDG} and requiring  $\chi^2<30$. The updated momentum of each photon from the kinematic fit is used in the further analysis.

\begin{figure*}[htbp!]
\centering
%\begin{overpic}
%[width=1\linewidth]{ST_all.eps}
\includegraphics[width=1\linewidth]{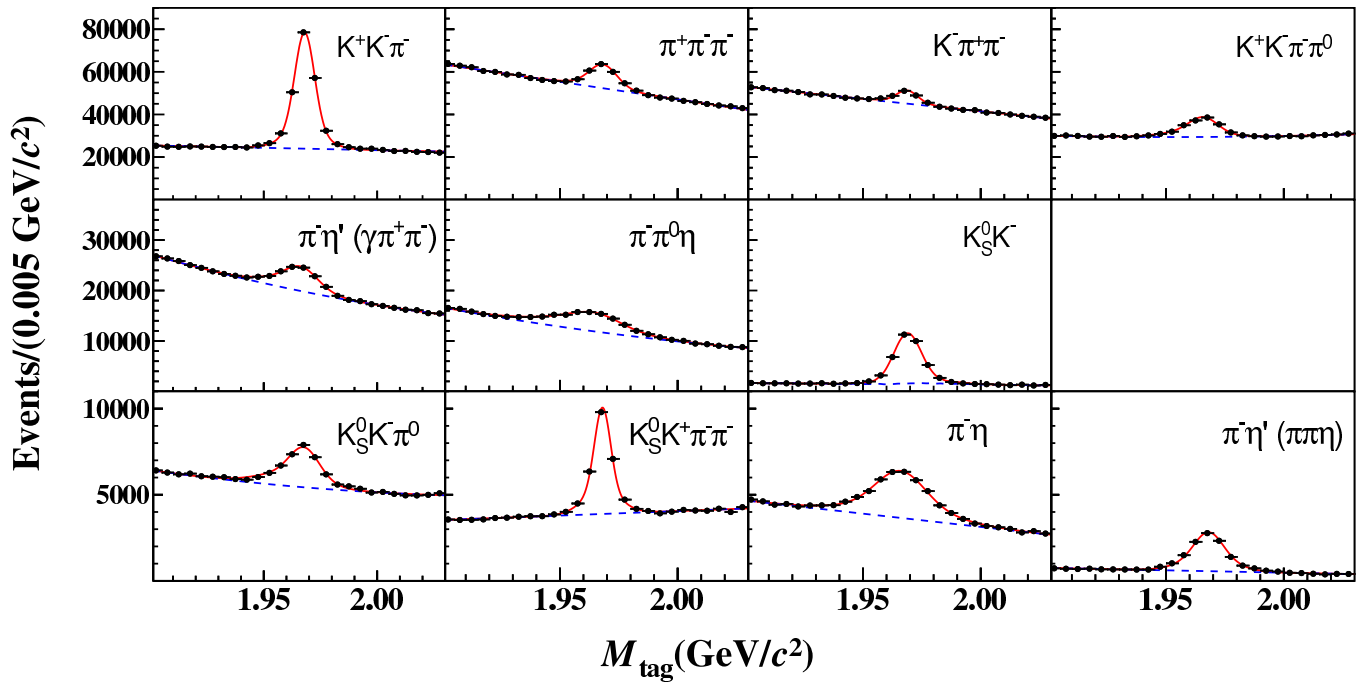}
%\end{overpic}

\caption{(Color online) Fits to the $M_{\rm tag}$ distributions for various ST modes.
The dots with error bars show data, the red solid lines are the overall fit results, and the blue dashed curves are the background.}
\label{ST_fit}
\end{figure*}

The $\ks$ candidates are reconstructed via the decay $\ks \to \pi^{+}\pi^{-}$ by performing a vertex-constrained fit to all oppositely charged track pairs without PID requirements applied.
The charged tracks must be within $|\cos\theta|<0.93$, and have a point of closest approach to the IP within $\pm$20\,cm along the beam direction; no requirement is placed on the point of closest approach in the plane perpendicular to the beam.
The $\chi^2$ of the vertex fit must be less than 100.
To suppress the combinatorial background, a secondary vertex fit is performed, constraining the direction of the $\ks$ momentum to point back to the IP, and requiring  $\chi^2<20$.
The flight length $L$, defined as the distance between the common vertex of the $\pi^+\pi^-$ pair and the IP, is obtained in the secondary vertex fit and required to satisfy $L>2\sigma_L$ for accepted $\ks$ candidates, where $\sigma_L$ is the uncertainty of $L$.
The four-momenta after the secondary vertex fit are used in the subsequent analysis.
The $\ks$ candidate is required to have a mass within the range (0.487, 0.511)\,$\gevcc$, corresponding to three standard deviations on the mass distribution.

The $\etap$ candidates are reconstructed via the prominent decay modes $\etap\to\pi^+\pi^-\eta$ and $\etap\to\gamma \pi^+\pi^-$, requiring the invariant masses of ${\pi^+\pi^-\eta}$ and ${\gamma\pi^+\pi^-}$ to be within (0.945, 0.970) and (0.938, 0.978)\,$\gevcc$, respectively. The $\rho^{\pm(0)}$ candidate is selected by requiring the $\pi^\pm\pi^{0(\mp)}$ invariant mass within (0.6, 0.9)\,$\gevcc$.

In the ST mode $\dsm\to\modek$, the $\pi^{+}\pi^{-}$ invariant mass is required to be outside the range (0.480, 0.515)\,$\gevcc$ to avoid double counting with the ST mode $\dsm\to\modea$.

For a given ST mode, the $\dsm$ candidates are reconstructed by all possible combinations of selected $K^\pm$, $\pi^\pm$, $\ks$, $\pi^0$, $\eta$ and $\etap$ candidates in an event, and are identified with the corresponding invariant mass $M_{\rm tag}$.
To suppress the background from the non-strangeness excited $D^*$ decay $D^*\to\pi D$, the $\pi^{\pm(0)}$ candidates from $\dsp$ decays must have a momentum larger than 100\,$\mevc$.
To further suppress the non $\dssp\dsm$ backgrounds,
a variable that represents the invariant mass of the system recoiling against the selected $\dsm$ candidate is defined as
\begin{linenomath*}
\begin{equation}
\mrec^2 = \left(\ecm-\sqrt{|\vec{p}_{D_s}|^{2}+M_{D_s}^2}\right)^2-|\vec{p}_{D_s}|^{2},
\end{equation}
\end{linenomath*}
where $\ecm$ is the center-of-mass energy, $\vec{p}_{\ds}$ is the momentum of the selected $\dsm$ candidate in the center-of-mass system, and $M_{\ds}$ is the nominal $\dsm$ mass~\cite{PDG}.
In the process $\ee\to\dssp\dsm\to\dsp\gamma\dsm$, the selected $\dsm$ candidates are produced either directly in the $\ee$ annihilation or from the decay $\dssm\to\gamma\dsm$.
The corresponding $\mrec$ distribution for the former case peaks at the nominal $\dssp$  mass $M_{\dssp}$~\cite{PDG} smeared by the mass resolution, and that for the latter case has a relatively flat distribution between 2.05 and 2.18\,$\gevcc$.
The $\dsm$ candidates are accepted by requiring $2.05<\mrec<2.18\,\gevcc$. The $\ee\to\dsp\dsm$ process is highly suppressed by this requirement.
For an event with multi-$\dsm$ candidates for a specific tag mode per charge, only the one with minimum $|\mrec-M_{\dssp}|$ is kept.

The $M_{\rm tag}$ distributions of the events passing the above selection criteria are shown in Fig.~\ref{ST_fit} for all ST modes. The ST yields are determined by performing a binned maximum likelihood fit.
In the fit, the $\dsm$ signal is described by the MC-simulated line-shape convolved with a Gaussian function representing the resolution difference between data and MC simulation, where the parameters of the Gaussian functions are free parameters the fit. The background is described by Chebychev polynomial functions of the first kind of first or second order.
The fit results are superimposed on the data in Fig.~\ref{ST_fit}.
For further study, we require that $M_{\rm tag}$ is within 2.5 times the resolution around the $\dsm$ peak.
The requirements on $M_{\rm tag}$, the ST yields, and the corresponding ST
detection efficiencies obtained with the generic MC samples are summarized in Table~\ref{ST_T} for each individual ST mode.

The signal $\dspn$ and the isolated photon from the $\dss$ decay are reconstructed from the remaining tracks and photons that are not used in the ST $\dsm$ reconstruction.
Exactly one remaining charged track with opposite charge to the ST $\dsm$ meson and at least one remaining good photon are required. The charged track is identified as a proton by requiring $\mathcal{L}(p)\geq \mathcal{L}(K)$, $\mathcal{L}(p)\geq \mathcal{L}(\pi)$ and $\mathcal{L}(p)\geq 0.001$.
The angle between this isolated photon and the nearest charged track is required to be larger than 10$^\circ$.

\begin{figure}[ht!]
\centering
\includegraphics[width=0.9\linewidth]{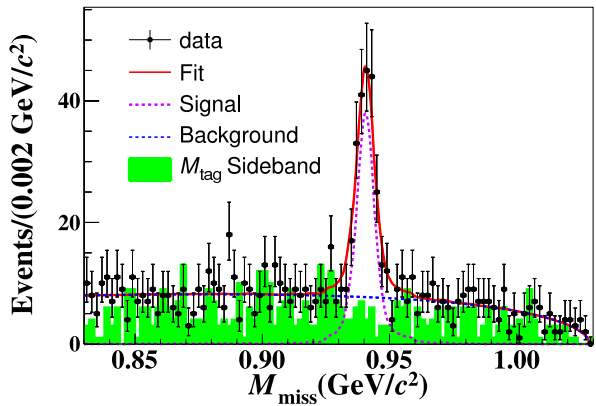}
\caption{(Color online) Fit to the $\mmiss$ distribution. The dots with error bars represent data, the (green) shaded histogram shows the events in the $M_{\rm tag}$ sideband region. The (red) solid line is the overall fit, the (violet) dotted line is the signal component, and the (blue) dashed line is the background component from the fit.}
\label{DT_fit}
\end{figure}

To improve the resolution and the likelihood of associating the correct photon candidate from the $D_s^*$ decay, we perform a kinematic fit with constraints on the masses of the ST $\dsm$, signal $\dsp$, intermediate state $D_s^{*\pm}$, and the initial four-momentum.
The two hypotheses, \emph{i.e.}, $\ee\to\dssp(\gamma+p\bar{n})\dsm(ST)$ or $\ee\to\dsp(p\bar{n})\dssm(\gamma+ST)$, are tested, and the one with the smaller fit $\chi^2$ is chosen.
In the fit, the antineutron is treated as a missing particle with unknown mass, thus there are a total of $7-4=3$ constraints.
The $\chi^2$ of the kinematic fit is required to be less than 200. This requirement retains most of the signal events, but removes 50\% of background.
For an event with more than one remaining photon, we try all photon candidates in the kinematic fit, and the one with the smallest $\chi^2$ is selected.
The updated momenta after the kinematic fit are used in the subsequent analysis. The resulting mass of the missing particle $\mmiss$, using all ST modes, is shown in Fig.~\ref{DT_fit}. A prominent antineutron signal is visible.

\begin{table}[htbp!]
\caption{Requirements on $M_{\rm tag}$, ST yields, ST and DT detection efficiencies for individual ST modes. The uncertainties are statistical only. The BFs of $\pi^0/\eta\to\gamma\gamma, \ks\to\pi^+\pi^-, \eta'\to \pi^+\pi^-\eta$ and $\eta'\to \gamma\pi^+\pi^-$ are not included in efficiencies.}{\label{ST_T}}
\begin{center}
%\footnotesize
\scriptsize
\begin{tabular}{lcccc}
\hline
\hline
ST mode                         & $M_{\rm tag}(\gevcc)$   & $N_{\rm ST}^i$ & $\epsilon_{\rm ST}^i$(\%)  & $\epsilon_{\rm DT}^i$(\%)  \\
\hline
$\ks K^-$                    & [1.950,1.990] & $30364\pm231$  & $46.23\pm0.04$  & $19.12\pm0.95$ \\
$K^+K^-\pi^-$                & [1.950,1.985] & $133666\pm544$ & $39.67\pm0.02$  & $17.85\pm0.40$ \\
$\ks K^-\pi^0$               & [1.930,1.990] & $10425\pm316$  & $15.45\pm0.03$  & $9.39\pm0.81$ \\
$K^+K^-\pi^-\pi^0$           & [1.930,1.990] & $37299\pm633$  & $10.46\pm0.01$  & $5.52\pm0.24$ \\
$\ks K^+\pi^-\pi^-$          & [1.950,1.985] & $13475\pm350$  & $18.74\pm0.03$  & $10.00\pm0.66$ \\
$\pi^+\pi^-\pi^-$            & [1.950,1.985] & $34918\pm688$  & $50.32\pm0.03$  & $23.08\pm1.07$ \\
$\pi^-\eta$                  & [1.930,2.000] & $16951\pm222$  & $42.83\pm0.04$  & $23.10\pm1.59$  \\
$\pi^-\pi^0\eta$             & [1.920,1.995] & $27631\pm785$  & $14.69\pm0.01$  & $9.04\pm0.55$ \\
$\pi^-\eta'(\pi^+\pi^-\eta)$ & [1.940,2.000] & $8675\pm120$   & $21.51\pm0.04$  & $8.98\pm0.78$ \\
$\pi^-\eta'(\gamma\pi^+\pi^-)$     & [1.945,1.980] & $22720\pm524$  & $27.48\pm0.03$  & $13.49\pm1.04$ \\
$K^-\pi^+\pi^-$              & [1.950,1.985] & $15801\pm463$  & $44.82\pm0.04$  & $23.64\pm1.75$ \\

\hline
\hline
\end{tabular}
\end{center}
\end{table}

The potential backgrounds are classified into (a) non-$D^-_s$ background and (b) real-$D^-_s$ background. The background (a) is dominated by continuum processes with proton and antineutron in the final state and can be estimated with the events in the $M_{\rm tag}$ sideband region (3.5$\sim$5.0$\sigma$ away from the $D_s$ peak). The corresponding $\mmiss$ distribution of background (a) is shown as the shaded histogram in Fig.~\ref{DT_fit}. No obvious peak is observed in the vicinity of the anti-neutron signal.
Since $\dspn$ is the only baryonic decay mode for the $\dsp$ meson, no peaking background is expected for background (b).
The properties of the backgrounds are validated by studying the generic MC samples.

The total DT signal yield
is determined by performing an unbinned maximum likelihood fit to the $\mmiss$ distribution in Fig.~\ref{DT_fit}, where the signal is described by an MC-simulated line-shape convolved with a Gaussian function representing the resolution difference between data and MC simulation; the background is modeled by an ARGUS function~\cite{argus}.
The fit shown in Fig.~\ref{DT_fit} returns $193\pm17$ $\dspn$ signal events. %corresponding  to a statistical significance of greater than 10$\sigma$.
The DT efficiencies for the individual ST mode are estimated by performing the same procedure on the generic MC samples, and are summarized in Table~\ref{ST_T}.
Based on Eq.~(\ref{equation3}), inserting all the numbers reported above and incorporating the world-average value for $\br{}(\dssp\to\gamma\dsp)$~\cite{PDG}, we obtain $\br{}(\dspn)=(1.21\pm0.10)\times10^{-3}$, where the uncertainty is statistical only.

With a DT technique, the systematic uncertainties on detecting the ST $\dsm$ meson largely cancel.
For the reconstruction of the isolated photon and the signal $\dspn$, the following sources of systematic uncertainties are studied, resulting in a total systematic uncertainty of 4.4\% when the individual contributions are summed in quadrature.

The efficiencies for proton tracking and PID are studied as function of $\cos\theta$ and momentum using the control sample $\ee\to \pi^+\pi^- p\bar{p}$. The results are then weighted by the $\cos\theta$ and momentum distributions of the proton in the signal MC. The average efficiency difference between data and MC simulation combined for tracking and PID is 3.2\%,
which is taken as the systematic uncertainty.

We study the uncertainties associated with the photon detection and the kinematic fit simultaneously with a control sample of $\dsp\to\ks K^+$ decays produced in the process $\ee\to \dssp\dsm\to \dsp\gamma\dsm$.
The resultant difference on the efficiencies between data and MC simulation is 2.4\%, which is assigned as the systematic uncertainty from this source.

The proton and antineutron may produce additional showers in the EMC that might then affect the efficiency of detecting $D^+_s\to p\bar{n}$ decays. To estimate this effect, we examine the detection efficiencies determined with two different signal MC samples that are produced with and without the neutron interaction effect in the EMC, respectively. Conservatively, we assign half of the difference between the two efficiencies, 0.9\%, as the uncertainty.

The uncertainty sources associated with the fit to the $\mmiss$ distribution include the background parameterization  and the fit range.
The corresponding uncertainties are estimated by performing fits with alternative background shape obtained with the events in the ST $M_{\rm tag}$ sideband region and various fit ranges.
The resultant changes on the signal yields are regarded as the corresponding uncertainties. The sum of the three uncertainties above in quadrature is 0.7\%, which is taken as the associated systematic uncertainty.

For the ST $\dsm$ yields, there is a contribution from the process $\ee\to\gamma_{\rm ISR}\dsp\dsm$, which causes a tail falling into the $\mrec$ windows.
We estimate this background contributes to our ST yields by at most 0.3\% based on the MC simulation. We take this upper limit as the systematic uncertainty from this source.

According to Eq.~(\ref{3}), the uncertainty related to the ST efficiency is expected to be canceled.
However, due to the different multiplicities,
the ST efficiencies estimated with the generic and the signal MC samples are expected to differ slightly.
Thus, the uncertainty associated with the ST efficiency is not canceled fully, which results in a so called ``tag bias'' uncertainty.
We study the tracking/PID efficiencies in different multiplicities, and take the combined differences between data and MC simulation,
0.6\%, as the corresponding uncertainty.

The uncertainties associated with the quoted BF of $\dssp\to\gamma\dsp$ and the limited MC statistics are also considered, which lead to 0.8\% and 1.1\%, respectively.

%\section{Summary}
In summary, using an $\ee$ collision data sample corresponding to 3.19~fb$^{-1}$ collected at  $\sqrt{s}=4.178\,\gev$ with the BESIII detector, we report the observation of $\dspn$ %with a statistical significance larger than 10 standard deviations
and measure the absolute BF to be $(1.21\pm0.10\pm0.05)\times10^{-3}$, where the first uncertainty is statistical and second systematic.
The decay $\dspn$ is confirmed and the precision of the BF measurement is much better than that of the previous measurement~\cite{Cleo-c:pn}. The anomalously large BF for $\dspn$ explicitly shows that the weak annihilation process featured as a short-distance dynamics is not the driving mechanism for this transition, while the hadronization process driven by non-perturbative dynamics determines the underlying physics. The measurement is important since similar annihilation effect is also present in other hadronic decays of charmed mesons. Relating this baryonic decay rate to the leptonic rate should provide important clues on how baryons are produced in hadronic interactions. The improved measurement also sets up the non-perturbative scale, allowing a better understanding of the transition mechanism.
This high precision measurement gives clear evidence for the role played by the hadronization process and is useful for improving existing and developing further models.

%\newpage
%%%%%%%%%%%%%%%%%%%%%%%%%%%%%%%%%%%%%%%%%%%%%%%%%%%%%%%%%%%%%%%%
%%%%%    acknowledgments       Part                %%%%%%%%%%%%%
%%%%%%%%%%%%%%%%%%%%%%%%%%%%%%%%%%%%%%%%%%%%%%%%%%%%%%%%%%%%%%%%
%\begin{acknowledgments}
The BESIII collaboration thanks the staff of BEPCII, the IHEP computing center and the supercomputing center of USTC for their strong support.
The authors are grateful to Prof. Hai-Yang Cheng, Dr. Xian-Wei Kang and Dr. Fu-Sheng Yu for enlightening discussions.
This work is supported in part by National Key Basic Research Program of China under Contract No. 2015CB856700; National Natural Science Foundation of China (NSFC) under Contracts Nos. 11405046, 11605198, 11235011, 11335008, 11425524, 11625523, 11635010, 11375170, 11275189, 11475164, 11475169, 11605196, 11705192; the Chinese Academy of Sciences (CAS) Large-Scale Scientific Facility Program; the CAS Center for Excellence in Particle Physics (CCEPP); Joint Large-Scale Scientific Facility Funds of the NSFC and CAS under Contracts Nos. U1332201, U1532102, U1532257, U1532258, U1732263, U1832103; CAS Key Research Program of Frontier Sciences under Contracts Nos. QYZDJ-SSW-SLH003, QYZDJ-SSW-SLH040; 100 Talents Program of CAS; National 1000 Talents Program of China; INPAC and Shanghai Key Laboratory for Particle Physics and Cosmology; German Research Foundation DFG under Contracts Nos. Collaborative Research Center CRC 1044, FOR 2359; Istituto Nazionale di Fisica Nucleare, Italy; Koninklijke Nederlandse Akademie van Wetenschappen (KNAW) under Contract No. 5304CDP03; Ministry of Development of Turkey under Contract No. DPT2006K-120470; National Natural Science Foundation of China (NSFC) under Contracts Nos. 11505034, 11575077; National Science and Technology fund; The Swedish Research Council; U. S. Department of Energy under Contracts Nos. DE-FG02-05ER41374, DE-SC-0010118, DE-SC-0010504, DE-SC-0012069; University of Groningen (RuG) and the Helmholtzzentrum fuer Schwerionenforschung GmbH~(GSI), Darmstadt; WCU Program of National Research Foundation of Korea under Contract No. R32-2008-000-10155-0.

%\section{References}


\begin{thebibliography}{99}

\bibitem{Gershtein:1976aq}
  S.~S.~Gershtein and M.~Y.~Khlopov,
  %``Lepton Decays of Heavy Pseudoscalar Meson,''
  Pisma Zh.\ Eksp.\ Teor.\ Fiz.\  {\bf 23}, 374 (1976).
  %%CITATION = ZFPRA,23,374;%%
  %15 citations counted in INSPIRE as of 11 Jan 2019

\bibitem{Pham:1980}
  X.~Y.~Pham, Phys.\ Rev.\ Lett. {\bf 45}, 1663 (1980).
\bibitem{Bediaga:1992}
  I.~Bediaga and E.Predazzi, Phys.\ Lett.\ B {\bf 275}, 161 (1992).
\bibitem{Chen:2008pf}
  C.~H.~Chen, H.~Y.~Cheng and Y.~K.~Hsiao,
  %``Baryonic D Decay D(s)+ ---> p anti-n and Its Implication,''
  Phys.\ Lett.\ B {\bf 663}, 326 (2008).
  %doi:10.1016/j.physletb.2008.04.033
  %[arXiv:0803.2910 [hep-ph]].
\bibitem{Bigi:1992}
I.\ I.\ Y.\ Bigi and N.\ G.\ Uraltsev,
Phys.\ Lett.\ B {\bf 280}, 271 (2008).

\bibitem{Cleo-c:pn}
S.~B.~Athar {\it et al.} [CLEO Collaboration],
  %``First Observation of the Decay D(s)+ ---> p anti-n,''
  Phys.\ Rev.\ Lett.\  {\bf 100}, 181802 (2008).
  %doi:10.1103/PhysRevLett.100.181802
  %[arXiv:0803.1118 [hep-ex]].
  %%CITATION = doi:10.1103/PhysRevLett.100.181802;%%
  %17 citations counted in INSPIRE as of 30 Aug 2016

\bibitem{Hsiao:2015}
 Y.~K.~Hsiao and C.~Q.~Geng, Phys.\ Rev.\ D {\bf 91}, 077501 (2015).

\bibitem{:2009vd}
  M.~Ablikim {\it et al.}  [BESIII Collaboration], Nucl.\ Instrum.\ Meth.\ A {\bf 614}, 345 (2010).

\bibitem{Wang:2016bzv}
  X.~Wang {\it et al.},
  %``The upgrade system of BESIII ETOF with MRPC technology,''
  JINST {\bf 11}, C08009 (2016).
  %%CITATION = doi:10.1088/1748-0221/11/08/C08009;%%
  %1 citations counted in INSPIRE as of 13 Aug 2017



\bibitem{geant4}
  S.~Agostinelli {\it et al.}  [GEANT4 Collaboration], Nucl.\ Instrum.\ Meth.\ A {\bf 506}, 250 (2003).

\bibitem{Ping:2013jka}
  R.~G.~Ping,
  %``An exclusive event generator for $e^+ e^-$ scan experiments,''
  Chin.\ Phys.\ C {\bf 38}, 083001 (2014).

\bibitem{photos}
  E.~Richter-Was, Phys.\ Lett.\ B {\bf 303}, 163 (1993).

\bibitem{evtgen}
  D.~J.~Lange,
  Nucl.\ Instrum.\ Meth.\ A {\bf 462}, 152 (2001);
  R.~G.~Ping,
  Chin. Phys. C {\bf 32}, 599 (2008).

\bibitem{PDG}
  C.~Patrignani {\it et al.} [Particle Data Group],
  %``Review of Particle Physics,''
  Chin.\ Phys.\ C {\bf 40}, 100001 (2016).

\bibitem{lund}
  J.~C.~Chen, G.~S.~Huang, X.~R.~Qi, D.~H.~Zhang and Y.~S.~Zhu, Phys.\ Rev.\ D {\bf 62}, 034003 (2000).


\bibitem{t0}
X.~Ma {\it et al.}, Chin. Phys. C {\bf 32}, 744 (2008);
Y.~H.~Guan, X.~R.~Lu, Y.~H.~Zheng and Y.~F.~Wang, Chin. Phys. C {\bf 38}, 016201 (2014).

\bibitem{argus} H. Albrecht {\it et al}. [ARGUS Collaboration], Phys. Lett. B {\bf 241}, 278 (1990).


\end{thebibliography}
\end{document}